\documentclass[nofootinbib,aps,preprintnumbers,unsortedaddress,prl,twocolumn]{revtex4}
\usepackage{graphicx}
\usepackage{cancel}
\usepackage{amssymb}
\usepackage{textcomp}
\usepackage{amsmath}
\usepackage{bm}
\usepackage{times}
\usepackage{epsfig}
\usepackage{color}
\usepackage{graphics}
\usepackage{hyperref}
\usepackage{setspace}
\usepackage{paralist}
\usepackage{epsfig}
\usepackage{color}
\usepackage{slashed}
\usepackage{comment}
\usepackage{epstopdf}
\usepackage{array}
\usepackage{tikz}

\usepackage{rotating}
\usepackage{framed}

\hypersetup{
    pdfnewwindow=true,      % links in new window
    colorlinks=true,       % false: boxed links; true: colored links
    linkcolor=black,          % color of internal links
    citecolor=blue,        % color of links to bibliography
    filecolor=blue,      % color of file links
    urlcolor=blue           % color of external links
}

\makeatletter
\newcommand\xleftrightarrow[2][]{%
  \ext@arrow 9999{\longleftrightarrowfill@}{#1}{#2}}
\newcommand\longleftrightarrowfill@{%
  \arrowfill@\leftarrow\relbar\rightarrow}
\makeatother

\begin{document}
\title{\Large {\it{\bf{Simple Left-Right Theory: Lepton Number Violation at the LHC}}}}
\author{Pavel Fileviez Perez$^{1,2}$}
\email{fileviez@mpi-hd.mpg.de}
\author{Clara Murgui$^{2}$}
\email{murgui@mpi-hd.mpg.de}
\author{Sebastian Ohmer$^{2}$}
\email{sebastian.ohmer@mpi-hd.mpg.de}
\affiliation{\vspace{0.15cm} \\  
$^{1}$ CERCA, Physics Department, Case Western Reserve University, 
Rockefeller Bldg. Cleveland, OH 44106, USA. \\
$^{2}$Particle and Astro-Particle Physics Division \\
Max-Planck-Institut fuer Kernphysik {\rm{(MPIK)}} \\
Saupfercheckweg 1, 69117 Heidelberg, Germany}
%
%%%%%%%%%%%%%%%%%%%%%%%%%%%%%%%%%%%%%%%%%%%%%%%%%%%%%%%%%%%%%%%%%%%%
\date{\today}
%%%%%%%%%%%%%%%%%%%%%%%%%%%%%%%%%%%%%%%%%%%%%%%%%%%%%%%%%%%%%%%%%%%%
%\vspace{1.0cm}
\begin{abstract}
We propose a simple left-right symmetric theory where the neutrino masses are generated at the quantum level.
In this context the neutrinos are Majorana fermions and the model has the minimal degrees of freedom in the scalar sector needed for symmetry breaking and mass generation.
We discuss the lepton number violating signatures with two charged leptons of different flavor and missing energy at the Large Hadron Collider in order to understand the testability of the theory.
\end{abstract}
%%%%%%%%%%%%%%%%%%%%%%%%%%%%%%%%%%%%%%%%%%%%%%%%%%%%%%%%%%%%%%%%%%%%
\maketitle
%%%%%%%%%%%%%%%%%%%%%%%

%%%%%%%%%%%%%%%%%%%
{\bf{Introduction}}:
%%%%%%%%%%%%%%%%%%%
The Standard Model of particle physics describes with a great precision most of the experiments in particle physics.
The main goal of the Large Hadron Collider (LHC) is to discover new physics beyond the Standard Model and maybe 
establish a new theory which can provide a more complete description of nature. There are many appealing extensions of the 
Standard Model with unique signatures at the LHC. The so-called Left-Right Symmetric Theories~\cite{Pati:1974yy,Mohapatra:1974gc,Senjanovic:1978ev,Senjanovic:1975rk} 
based on the gauge symmetry, $SU(2)_L \otimes SU(2)_R \otimes U(1)_{B-L}$, have been considered for a long time as one of the most attractive theories for physics beyond the Standard Model. 

In the context of Left-Right Symmetric Theories~\cite{Pati:1974yy,Mohapatra:1974gc,Senjanovic:1978ev,Senjanovic:1975rk}, one understands the origin of the $V-A$ interactions present in the Standard Model, 
the neutrinos are massive and one has a deep connection between 
spontaneous parity violation and the origin of neutrino masses~\cite{TypeI-LR}. 
In this context, the smallness of the neutrino masses is understood through the seesaw mechanism~\cite{TypeI,TypeI-LR}, where the right-handed neutrino masses are defined by 
the $SU(2)_R \otimes U(1)_{B-L}$ breaking scale. These theories also predict the possibility 
to observe lepton number violating processes at colliders as discussed in Ref.~\cite{Keung:1983uu}, which will be crucial to establish the Majorana nature of neutrinos.   
Finally, one can say that these theories define the path to grand unified theories based on the $SO(10)$ gauge symmetries.

In this letter, we revisit the left-right symmetric theories and propose a new mechanism to generate 
Majorana neutrino masses in this context. We show that using the minimal Higgs sector composed of two Higgs doublets, $H_L$ and $H_R$, needed to break the left-right symmetry and adding only one additional charged scalar singlet field we can generate Majorana neutrino masses at the quantum level through the Zee mechanism~\cite{Zee:1980ai}. We discuss the main features of this model. 

In order to understand the testability of this model, we investigate the lepton number violating signatures at the LHC.
In this context, the charged scalar singlet can be produced at the LHC through gauge interactions and since it decays mainly into 
two leptons one can have signatures with two charged leptons with different flavour and missing energy, $e_i^+ e_j^- E_T^{miss}$.
This simple theory can be considered as an appealing extension of the Standard Model.
\vspace{0.5cm}
\\
%%%%%%%%%%%%%%%%%%%%%%%%%%%%%%
{\bf{Left-Right Symmetric Theories}}:
%%%%%%%%%%%%%%%%%%%%%%%%%%%%%
The left-right symmetric theories~\cite{Pati:1974yy,Mohapatra:1974gc,Senjanovic:1978ev,Senjanovic:1975rk} are based on the gauge group
\begin{align*}
SU(2)_L \otimes SU(2)_R \otimes U(1)_{B-L}\,,
\end{align*}
where we omit the QCD gauge group for simplicity. In these theories, the fermions live in the representations
\begin{align*}
&Q_L = \begin{pmatrix} u_L \\ d_L \end{pmatrix} \sim (2, 1, 1/3) \,, \quad Q_R = \begin{pmatrix} u_R \\ d_R \end{pmatrix} \sim (1, 2, 1/3) \,,  \\
&\ell_L = \begin{pmatrix} \nu_L \\ e_L \end{pmatrix} \sim (2, 1, -1) \,, \quad \ell_R = \begin{pmatrix} \nu_R \\ e_R \end{pmatrix} \sim (1, 2, -1) \,.
\end{align*}
A bi-doublet Higgs which in our notation reads as
\begin{align*}
&\Phi =  \begin{pmatrix} \phi_1^0 && \phi_2^+ \\ \phi_1^- && \phi_2^0 \end{pmatrix}  \sim (2, 2, 0)\,,
\end{align*}
acquires a vacuum expectation value and generates masses for the quarks and charged leptons.
The relevant Yukawa interactions to generate masses for the charged fermions are given by
\begin{align*}
- \mathcal{L} \supset & \,  \bar{Q}_L \left( Y_1 \Phi + Y_2  \tilde{\Phi} \right) Q_R +  \bar{\ell}_L \left( Y_3  \Phi + Y_4  \tilde{\Phi} \right) \ell_R + \textrm{h.c.}\,, 
\end{align*}
where $\tilde{\Phi}= \sigma_2 \Phi^*  \sigma_2$. After symmetry breaking the charged fermion masses read as
\begin{eqnarray}
M_U &=& Y_1 v_1 + Y_2 v_2^*\,, \\
M_D &=& Y_1 v_2 + Y_2 v_1^*\,, \\
M_E &=& Y_3 v_2 + Y_4 v_1^*\,.
\end{eqnarray}
Here, $v_1$ and $v_2$ are the vacuum expectation values for the fields $\phi_1^0$ and $\phi_2^0$, respectively. 

In this context, one can generate Dirac neutrino masses and the neutrino mass matrix is given by
\begin{eqnarray}
M_\nu^D &=& Y_3 v_1 + Y_4 v_2^*\,.
\end{eqnarray}
Typically, in the context of left-right symmetric theories one investigates mainly the generation of Majorana neutrino masses. 
However, it is important to mention that when $v_2 \ll v_1$ and $Y_3 \ll Y_4$ one can have small Dirac neutrino masses without fine-tuning. 
In this scenario, the charged lepton and neutrino masses can be written as
\begin{eqnarray}
M_E &\approx & Y_4 v_1^*\,, \\
M_\nu^D &=& v_1 \left( Y_3 + M_E  \frac{v_2^*}{|v_1|^2} \right)\,. 
\end{eqnarray}
One needs extra Higgses to break $SU(2)_R \otimes U(1)_{B-L}$ to $U(1)_Y$ and recover the Standard Model gauge symmetry. The simplest extra Higgs sector is composed of two Higgs doublets $H_L \sim (2,1,1)$ and $H_R \sim (1,2,1)$~\cite{Senjanovic:1975rk}. It is important to stress that this is the minimal left-right theory consistent with all experimental constraints and predicts Dirac neutrinos. See also Ref.~\cite{Branco} for a discussion on Dirac neutrinos in left-right theories.
\vspace{0.5cm}
\\
%%%%%%%%%%%%%%%%%%%%%%%%%%%%%%
{\bf{Theoretical Framework}}:
%%%%%%%%%%%%%%%%%%%%%%%%%
The neutrinos could be Majorana fermions. Let us discuss the different possibilities to generate Majorana neutrino masses in the context of left-right symmetric theories. 
One can have different implementations of the seesaw mechanism in this context. Here, we list the simplest scenarios:
\begin{itemize}
\item { \textit{Type I~\cite{TypeI,TypeI-LR} plus Type II~\cite{TypeII} seesaw}}: The scalar sector is composed of the bi-doublet Higgs and two Higgs triplets $\Delta_L \sim (3,1,2)$ and $\Delta_R \sim (1,3,2)$. The triplet $\Delta_R$ breaks 
$SU(2)_R \otimes U(1)_{B-L}$ to $U(1)_Y$ generating masses for the right-handed neutrinos and through the canonical 
Type I seesaw generates masses for the left-handed neutrinos. Here, one cannot avoid the Type II contribution once $\Delta_L$ acquires a vacuum expectation value.

\item {\textit{Type III~\cite{TypeIII} seesaw}}: In this scenario, one adds fermionic triplets $\rho_L \sim (3,1,0)$ and $\rho_R \sim (1,3,0)$. In order to break $SU(2)_R \otimes U(1)_{B-L}$ to $U(1)_Y$ 
and generate neutrino masses, one needs the Higgs doublets $H_L \sim (2,1,1)$ and $H_R \sim (1,2,1)$. See Ref.~\cite{FileviezPerez:2008sr} for the implementation of this mechanism in left-right symmetric theories.
The fermionic triplets of Type III seesaw can be used to define an alternative left-right theory where the baryon and lepton numbers are independent gauge symmetries~\cite{Duerr:2013opa}.

\item {\textit{Radiative Seesaw Mechanism}}: In this article, we will propose the possibility to generate Majorana neutrino masses at the quantum level following the idea of the Zee mechanism~\cite{Zee:1980ai}. We show that
adding a charged scalar boson $\delta^+ \sim (1,1,2)$ and the Higgs doublets $H_L$ and $H_R$ one can generate neutrino masses in a consistent way.  This possibility was mentioned in Ref.~\cite{Pal:1990my}, but the results in this paper are not correct. Here, we will discuss how one can implement this mechanism in a realistic way. We would like to mention that the generic Zee model where both Higgs doublets couple to the leptons is allowed by the experimental values of the mixing angles and squared mass differences, see Ref.~\cite{He:2003ih} for a detailed discussion. See also Ref.~\cite{He:2011hs} for the study of the contrains coming from 
lepton number violating decays.
\end{itemize}
In order to generate neutrino masses at the quantum level, we add the charged scalar $\delta^+$ and use the following interactions
\begin{eqnarray}
- \mathcal{L} &\supset &  \lambda_L \ell_L  \ell_L  \delta^{+}   +  \lambda_R  \ell_R \ell_R  \delta^+  \nonumber \\
&+&   \lambda_1 H^T_L i \sigma_2 \Phi H_R \delta^-  +  \lambda_2 H^T_L i \sigma_2 \tilde{\Phi} H_R \delta^-  + \text{h.c.} 
\end{eqnarray}
Notice that we will assume the discrete left-right parity only in the gauge sector since we are mainly interested in the case where the left-right scale is low.
It is well-known that the spontaneous breaking of a discrete symmetry leads to the domain wall problem~\cite{Zeldovich:1974uw}. We assume that this symmetry is explicitly 
broken by the Yukawa interactions and scalar potential. 
\begin{figure}[h]
\includegraphics[width=0.65\linewidth]{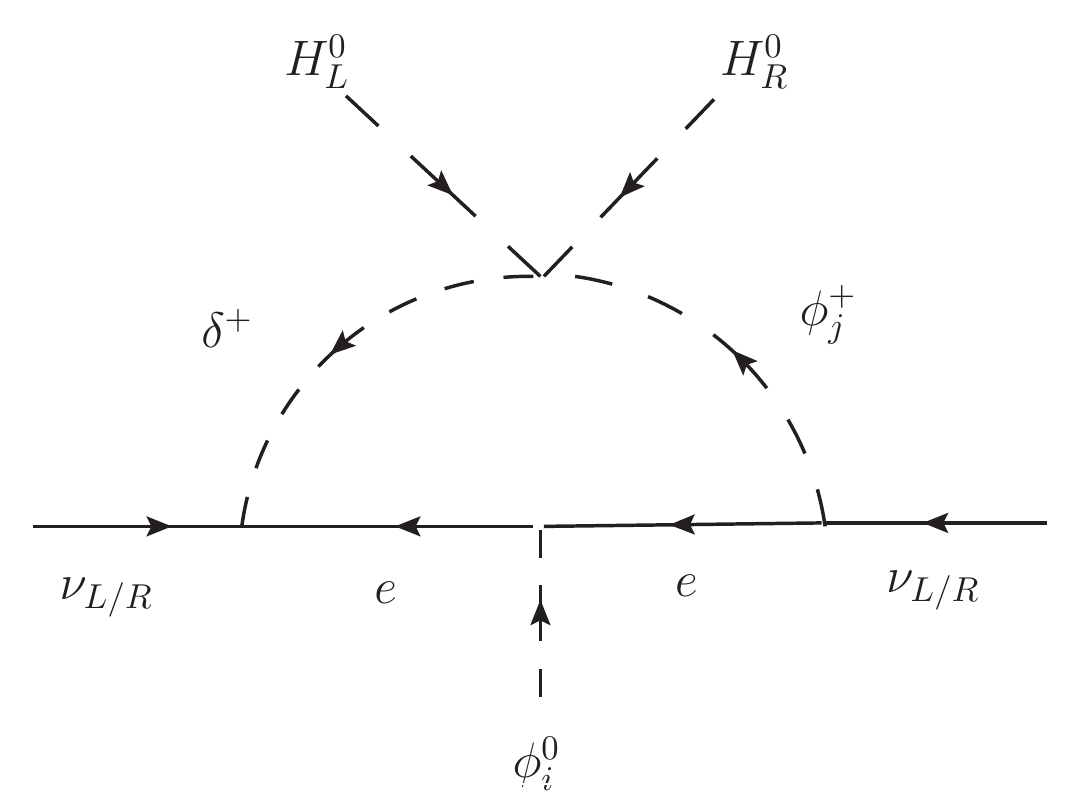}
\caption{Neutrino mass generation at the quantum level. This graph is shown in the unbroken phase only for ilustration.}
\label{Figunbroken}
\end{figure}

In order to compute the neutrino masses at the quantum level in the broken phase, we work in the physical basis where we find five charged scalar fields.  
In our notation, the mass matrix for the charged scalar fields is diagonalized by $V$,
\begin{align*}
\begin{pmatrix} \phi_1^+ \\ \phi_2^+ \\ h_L^+ \\ h_R^+ \\ \delta^+ \end{pmatrix} = V \begin{pmatrix} h_1^+ \\ h_2^+ \\ h_3^+ \\ h_4^+ \\ h_5^+ \end{pmatrix}.
\end{align*}
In Fig.~1, one can see that the simultaneous presence of the Yukawa and scalar quartic interaction for $\delta^+$ define the breaking of the total lepton number. 
After performing the loop calculation the neutrino mass matrix reads as
\begin{align*}
\begin{pmatrix} \nu && \nu^c \end{pmatrix}_L \begin{pmatrix} M_\nu^L && M_\nu^D \\ (M_\nu^D)^T && M_\nu^R \end{pmatrix}\begin{pmatrix} \nu \\  \nu^c \end{pmatrix}_L \,,
\end{align*}
where
\begin{eqnarray}
\displaystyle {(M_\nu^L)}^{\alpha \gamma}&=&\frac{1}{4\pi^2}\lambda_L^{\alpha \beta}m_{e_\beta}\sum_i \text{Log}\left(\frac{M_{h_i}^2}{m_{e_\beta}^2}\right) \times \nonumber \\
&& V_{5i}\left [ (Y_3^\dagger)^{\beta \gamma}V_{2i}^*-(Y_4^\dagger)^{\beta \gamma}V_{1i}^* \right ] \ + \ \alpha \leftrightarrow \gamma 
\,,\\
\displaystyle {(M^R_\nu)}^{\alpha \gamma}&=&\frac{1}{4\pi^2}\lambda_R^{\alpha \beta}m_{e_\beta}\sum_i\text{Log}\left(\frac{M_{h_i}^2}{m_{e_\beta}^2}\right) \times \nonumber \\
&& V_{5i}\left[(Y_3)^{\beta \gamma}V_{1i}^*-(Y_4)^{\beta \gamma}V_{2i}^*\right] \ + \ \alpha \leftrightarrow \gamma \,. 
\label{RHM}
\end{eqnarray}
Notice that in the above equations the two mass entries are proportional to the Yukawa coupling $\lambda_L (\lambda_R)$ and $V_{5i}$. 
These two entries have the information of the violation of the total lepton number. $M_\nu^D$ is given by Eq.(4) and in the limit when $v_2 \ll v_1$ and $Y_3 \ll Y_4$ it is given by Eq.(6).

In this model, we can have different scenarios for neutrino masses:
\begin{itemize}

\item When $\lambda_L \ll \lambda_R$ and $M_\nu^D \ll M^R_\nu$ one can have a low scale seesaw since the right-handed neutrinos will be light. We have discussed above that $M_\nu^D$ can be very small when $v_2 \ll v_1$ and $Y_3 \ll Y_4$. 
One can imagine a different case where $M_\nu^D$ is very small when $Y_4 \approx - Y_3 v_1/v_2^*$ and the other entries in the mass matrix are proportial to $Y_3$.

\item In the case when $M_\nu^L$ and $M^R_\nu$ are small compared to $M_\nu^D$, one can have quasi-Dirac neutrinos.

\end{itemize}

To estimate the mass scale of the Majorana mass of the right-handed neutrinos we assume one generation for simplicity and choose a conservative scenario with $Y_3$, $Y_4$, $\lambda_R$ $\sim$ $1$ and $m_\tau = 1.78$ GeV contributing to the loop factor. The sum over the logarithm of the charged scalar masses weighted by the combination of mixing matrices and Yukawa couplings can be approximated by twice the logarithm of the largest difference of the charged scalar masses due to the unitarity of the mixing matrices. If we assume that the lightest charged scalar field has a mass of order $100$ GeV and the heaviest charged scalar field is at the TeV scale we find $M_{\nu}^R \approx  0.4 \ \text{GeV}$ which shows how light the right-handed neutrinos are in this model under these assumptions.
\vspace{0.5cm}
\\
%%%%%%%%%%%%%%%%%%%
{\bf{Gauge Bosons}}:
%%%%%%%%%%%%%%%%%%%
As in any left-right symmetric theory, one has extra gauge bosons, the charged gauge bosons $W_R^{\pm}$ and a neutral gauge boson $Z^{'}$.
The predictions for the gauge boson masses in this model are identical to the model studied in Ref.~\cite{Senjanovic:1978ev}.
In the basis $(W_L^+  \ W_R^+)$ the charged gauge boson mass matrix reads as
\begin{equation}
\begin{pmatrix} \frac{g_L^2}{2}(\frac{1}{2}v_L^2 + v^2) && - g_L g_R v_1 v_2 \\ - g_L g_R v_1 v_2 && \frac{g_R^2}{2}(\frac{1}{2}v_R^2 + v^2) \end{pmatrix} \,, \nonumber
\end{equation}
where $v^2 = v_1^2 + v_2^2$. The mixing angle between $W_R$ and $W_L$ can be defined as $\tan 2 \theta_{LR} \approx 8 \frac{g_L}{g_R} \epsilon_{12}$, where $\epsilon_{12}=v_1 v_2 / v_R^2$. 

The mass matrix for the neutral gauge bosons in the basis $(W_L^0 \ W_R^0 \ Z_{BL})$ can be written as
\begin{equation}
 \begin{pmatrix} 
\frac{g_{L}^2}{2} (\frac{1}{2} v_L^2 + v^2) && - \frac{g_L  g_{R}}{2} v^2 && - \frac{g_L g_{BL}}{4} v_L^2 \\ 
- \frac{g_L g_{R}}{2} v^2 && \frac{g_R^2}{2} (\frac{1}{2}v_R^2 + v^2) && - \frac{g_R g_{BL}}{4} v_R^2  \\ 
- \frac{g_L g_{BL}}{4} v_L^2 && - \frac{g_R g_{BL}}{4} v_R^2 && \frac{g_{BL}^2}{4} (v_R^2+v_L^2) \end{pmatrix}\,. \nonumber
\end{equation}
In the limit $v_L \to 0$, one can make the following rotation, 
\begin{eqnarray}
W_L^0 &=& \cos \theta_W Z_L+\sin \theta_W A\,, \nonumber \\
W_R^0 &=& \cos \theta_R Z_R -\sin \theta_W \sin \theta_R Z_L + \cos \theta_W \sin \theta_R A\,, \nonumber \\
Z_{BL}^0 &=& -\sin \theta_R Z_R- \sin \theta_W \cos \theta_R Z_L + \cos \theta_W \cos \theta_R A\,, \nonumber 
\end{eqnarray}
which decouples the photon. Here, $\tan \theta_R = g_{BL}/g_R$ and $\tan \theta_W=g_Y/g_L$ with $g_Y = g_{BL}g_R/\sqrt{g^2_{BL}+g_R^2}$. 
After the photon decouples, one finds the mass matrix in the $(Z_R \ Z_L)$ basis
\begin{equation}
{\cal{M}}_{Z-Z^{'}}^2=  \begin{pmatrix} 
M_{RR}^2&& M_{LR}^2 \\ 
M_{LR}^2 &&  M_{LL}^2
\end{pmatrix}.
\end{equation}
where
\begin{eqnarray}
M_{RR}^2 &=& \frac{v_R^2}{4}\left(g_{BL}^2+g_R^2 \right) +\epsilon \ \frac{g_R^4 v_R^2}{2(g_{BL}^2+g_R^2)}\,, \\
M_{LR}^2&=& -\epsilon \ \frac{g_R^2 v_R^2\sqrt{g_L^2g_R^2 + g^2_{BL}(g_L^2+g_R^2)}}{2(g_{BL}^2+g_R^2)}\,, \\
M_{LL}^2 &=& \epsilon \  v_R^2 \ \frac{g_L^2g_R^2 + g^2_{BL}(g_L^2+g_R^2)}{2(g_{BL}^2+g_R^2)}\,,
\end{eqnarray}
with $\epsilon=(v_1^2 +v_2^2)/v_R^2$. Therefore, the mixing angle between the $Z$ and the heavy $Z^{'}$ reads as
\begin{equation}
\tan 2 \xi \approx \epsilon \ \frac{-4g_R^2 \sqrt{g_L^2g_R^2 + g^2_{BL}(g_L^2+g_R^2)}}{(g_{BL}^2+g_R^2)^2}\,,
\end{equation}
which has to be smaller than $10^{-3}$ to satisfy the electroweak precision constrains. If $v_R^2 \gg v_1^2 +v_2^2$ the bound is easily satisfied.
In the limit $v_R \gg v_1, v_2, v_L$, the masses of the new gauge bosons are given by
\begin{eqnarray}
M_{W_R} &\simeq & \frac{1}{2} g_R v_R \,, \\
M_{Z^{'}} &\simeq & \frac{\sqrt{g^2_{BL} + g^2_R}}{g_R} M_{W_R} \stackrel{g_L = g_R}{\simeq} 1.2 \, M_{W_R}\,.
\end{eqnarray}
Using the limit $M_{W_R} > 3$ TeV~\cite{Zhang:2007da} one can find a lower bound on the $Z^{'}$ mass, i.e. $M_{Z^{'}}>3.6$ TeV.
In this model the right handed neutrinos are light and they modify the leptonic branching ratio of the gauge bosons since one can have 
the decays $W_R^{+} \to \bar{e}_R \nu_R$ and $Z^{'} \to \bar{\nu}_R \nu_R$.

As we have discussed, the Higgs sector of this model is quite simple and one generates neutrino masses at the quantum level using the additional charged scalar $\delta^{+}$. 
This charged scalar field can give rise to lepton number violating interactions which can give rise to exotic processes. Here, we study one of the most exotic signatures in order 
to understand the testability of the model at the LHC.
\begin{figure}[h]
	\centering
		\includegraphics[width=0.5\textwidth]{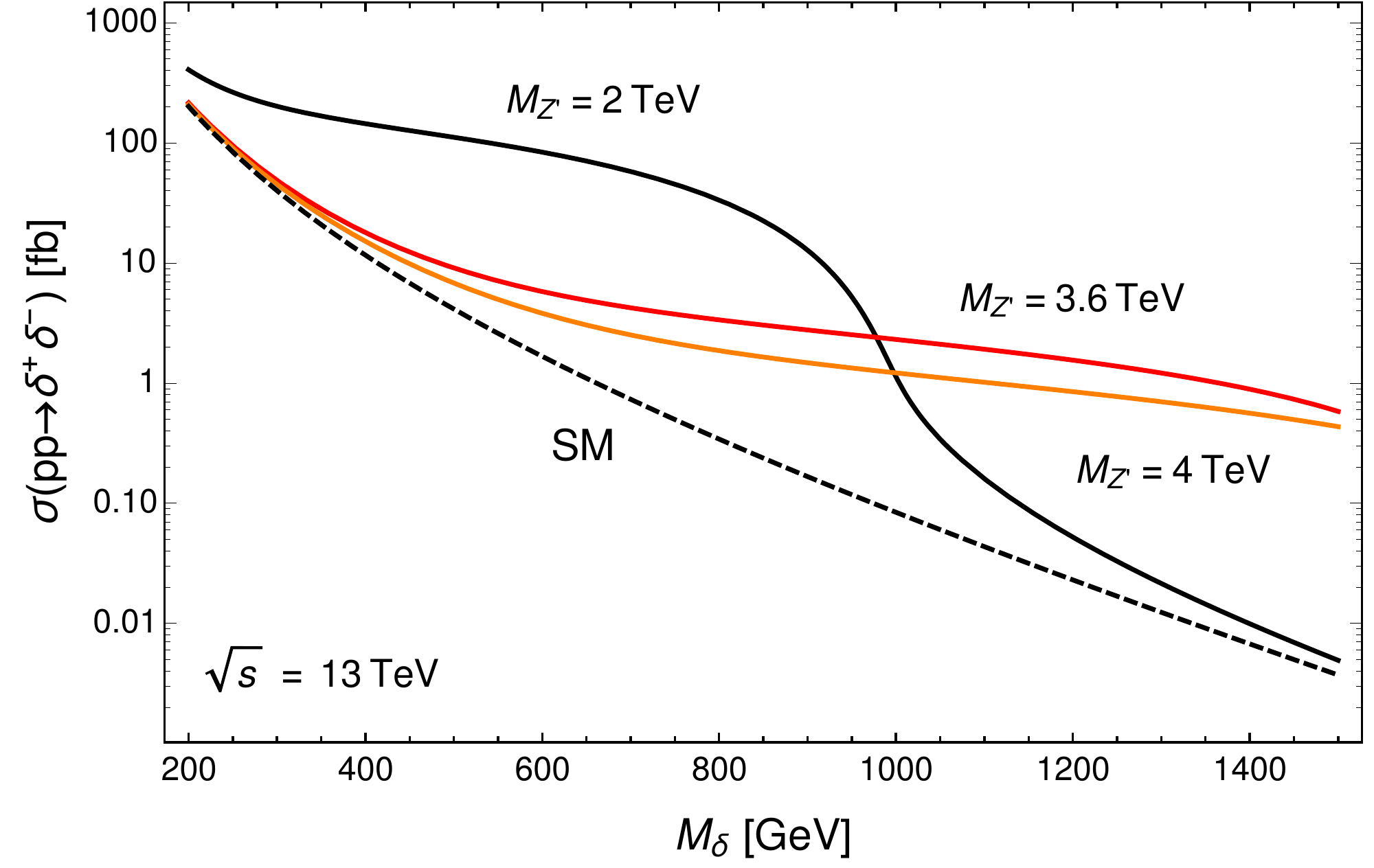}
		\caption{Production cross section of $\delta^+\delta^-$ as a function of $M_\delta$ for different $M_{Z^{'}}$ at $\sqrt{s} = 13$ TeV.}
		\label{fig_Delta_Prod}
\end{figure}
\vspace{0.5cm}
\\
%%%%%%%%%%%%%%%%%%%%%%%%%
{\bf{Lepton Number Violation at the LHC}}:
%%%%%%%%%%%%%%%%%%%%%%%%%
The extra charged scalar singlet $\delta^{\pm}$ couples to the leptons and gives rise to lepton number violating signals at the LHC.
The $\delta^{\pm}$ are produced through the Drell-Yan process, $$pp \to \gamma, Z, Z^{'} \to \delta^+ \delta^- \to e^{+}_i e^{-}_j E_T^{miss},$$ and one has signatures with two leptons of different flavors and missing energy.
The number of events for these channels is given by
\begin{eqnarray}
N (e^+_i e^-_j E_T^{miss}) &=& {\cal{L}} \times \sigma (pp \to \delta^+ \delta^-) \times \nonumber \\
&& {\rm{BR}} ( \delta^+ \to e^+_i \nu) \times {\rm{BR}} ( \delta^- \to \bar{\nu} e^{-}_j )\,, \nonumber 
\end{eqnarray} 
where $\cal{L}$ is the luminosity and $\sigma (pp \to \delta^+ \delta^-)$ is the cross section.  
In Fig.~2 we show the cross section versus the $\delta^{\pm}$ mass for different values 
of the $Z^{'}$ mass. The dashed line shows the prediction for the Standard Model contribution.
Here, the presence of the $Z^{'}$ allows us to have a larger cross section, see Fig.~2.
Notice that when the $\delta^{\pm}$ mass is below $650$ GeV the production cross section is above $1$ fb.
Here we neglect the mixing angle in the charged scalar sector for simplicity.

The charged scalar singlet $\delta^{\pm}$ in this model decays mainly into leptons when the mixing angle with the other charged scalars living 
in the bi-doublet is small. The only interaction which induces the coupling of $\delta^{\pm}$ to quarks is $H^T_L i \sigma_2 \Phi_i H_R \delta^-$.
Since the neutrino masses are also generated through this coupling as we showed in Fig.~1 this interaction will be small.  Therefore, 
the decay of the charged scalar singlet to quarks is generically suppressed. In our study, we assume for simplicity that $\delta^{\pm}$ decays mainly into leptons 
and discuss the lepton number violating signals at the LHC. 

\begin{figure}[h]
	\centering
		\includegraphics[width=0.5\textwidth]{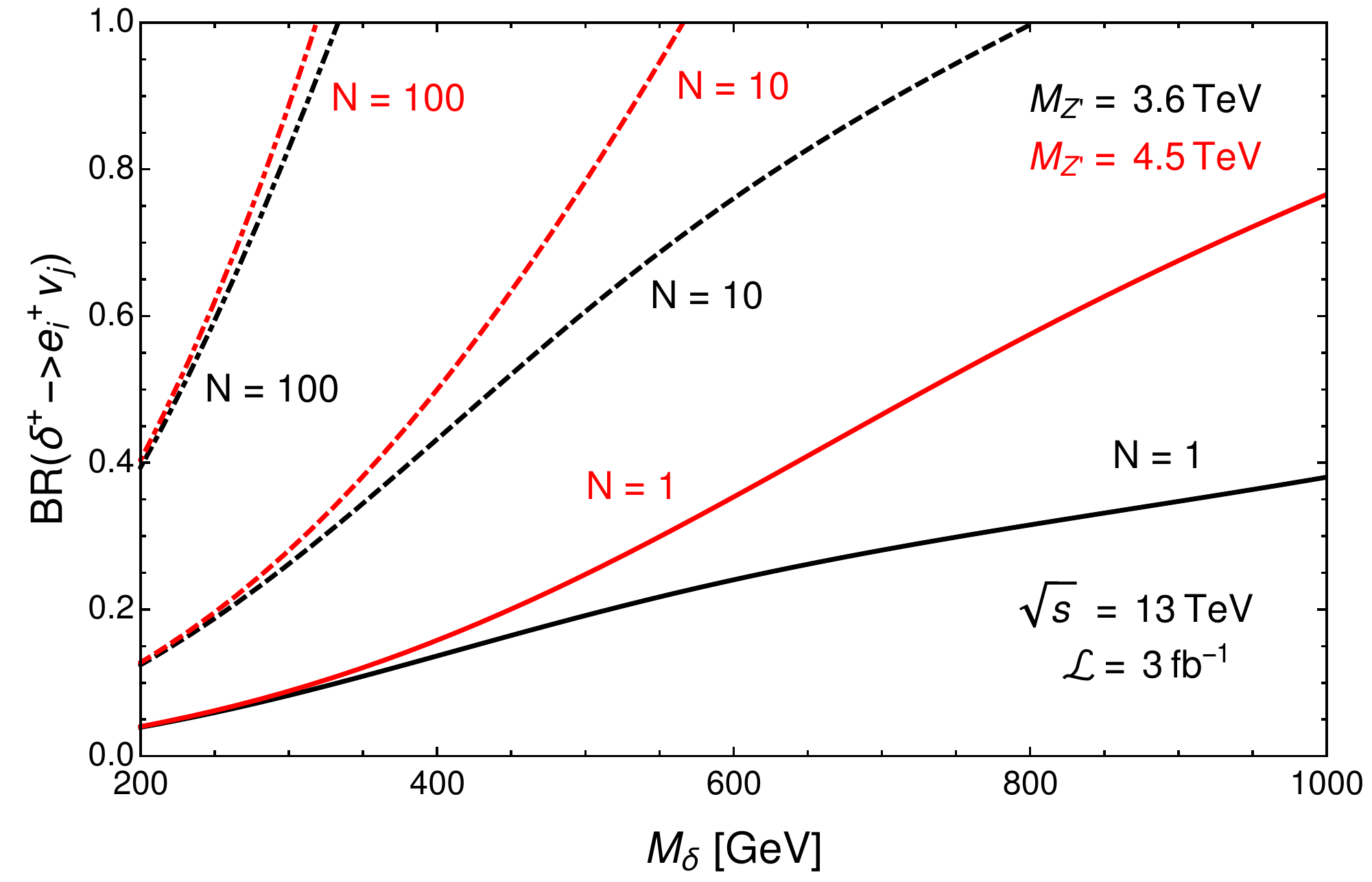}
\caption{Contours of expected number of events for $3 \ \rm{fb}^{-1}$ at $\sqrt{s} = 13$ TeV in the $M_\delta$--$\text{BR}(\delta^+ \to e_i^+ \nu_j)$ plane.}
		\label{fig_Delta_Prod}
\end{figure}		
		
\begin{figure}[h]
	\centering		
		\includegraphics[width=0.5\textwidth]{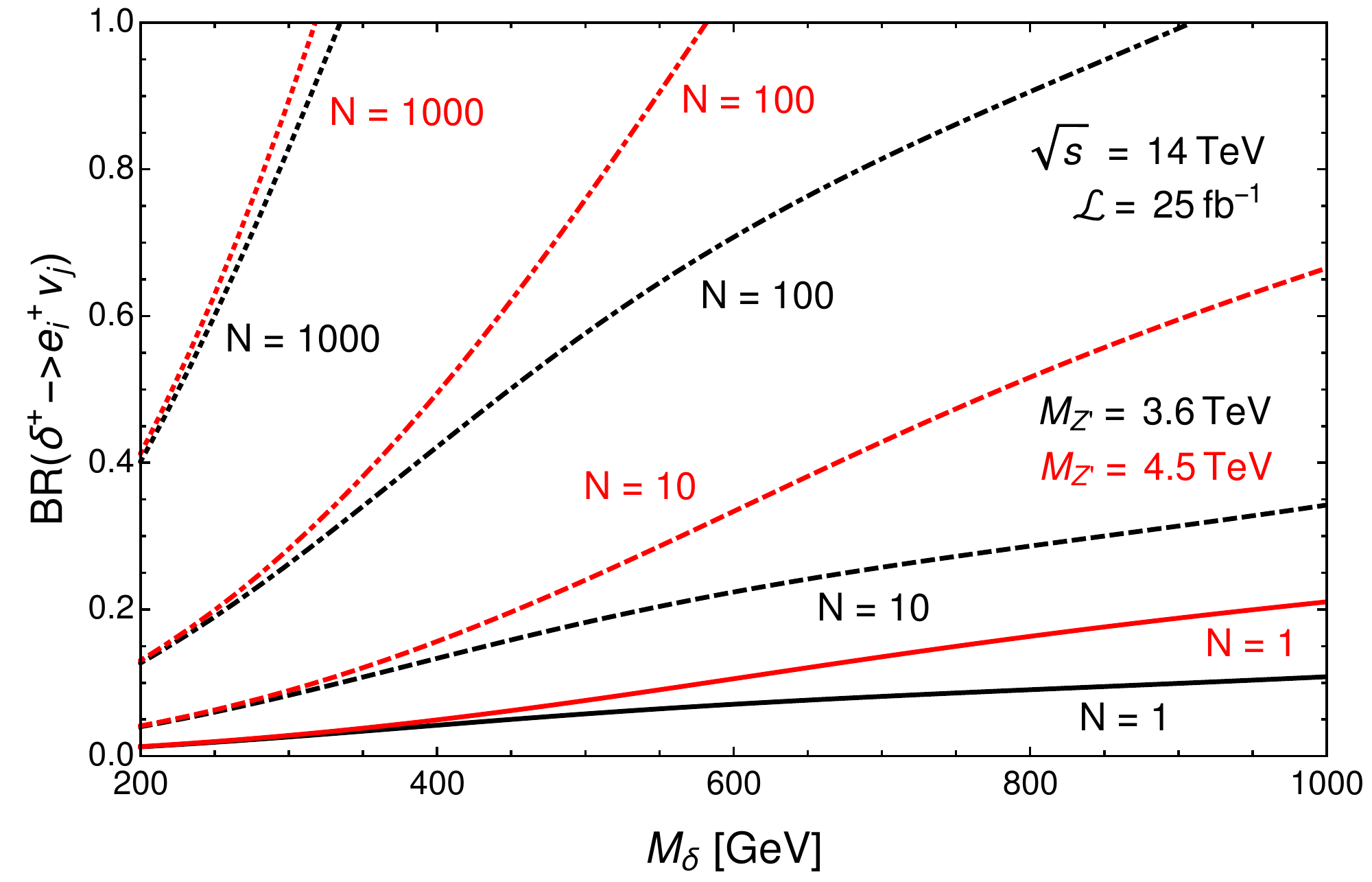}
		\caption{Contours of expected number of events for $25 \ \rm{fb}^{-1}$ at $\sqrt{s} = 14$ TeV in $M_\delta$--$\text{BR}(\delta^+ \to e_i^+ \nu_j)$ plane.}
		\label{fig_Delta_Prod}
\end{figure}

In Fig.~3 we show the isoplots for the number of events assuming a luminosity of ${\cal{L}}=3 \ \rm{fb}^{-1}$, $\sqrt{s}=13$ TeV and two values of the $Z^{'}$ mass.
Notice that when $M_{\delta}$ is below $500$ GeV and the ${\rm{BR}}(\delta^+ \to e^+_i \nu_j)$ is above $0.6$ one expects more than $10$ events.  
In Fig.~4 we show the predictions for the number of events when $\sqrt{s}=14$ TeV and luminosity of ${\cal{L}}=25 \ \rm{fb}^{-1}$. In this case, one can 
have a large number of events even when the $\delta^{\pm}$ mass is close to $1$ TeV.  The main backgrounds for these processes at the LHC are the $WW$ and $ZZ$ production. The $ZZ$ channel can never give rise to a channel with two leptons of different flavors. The $WW$ pair production can fake 
the signatures we study in this letter, but imposing a cut on the charged lepton transverse momentum can distinguish the signals. A detailed study of 
these events will be conducted in the near future. As one can appreciate, it will be easy to look for these signatures at the LHC and one could 
test this model in the near future.   
\vspace{0.5cm}
\\
%%%%%%%%%%%%%%%%%%%%
{\bf{Summary}}:
%%%%%%%%%%%%%%%%%%%%
We have proposed a simple left-right symmetric theory where the neutrinos are Majorana fermions and their masses are generated at the quantum level. 
In this model one has the simplest Higgs sector needed to break the left-right symmetry and generate Majorana neutrino masses. The Higgs sector is composed 
of two Higgs doublets $H_L$ and $H_R$, and an $SU(2)$ singlet $\delta^{\pm}$. This model has less degrees of freedom in the scalar sector 
than the canonical left-right symmetric theory with Higgs triplets, $\Delta_L$ and $\Delta_R$, where the neutrino masses are generated at tree level. 

We have investigated the main features of this model and discussed the possibility to observe lepton number violation at the LHC. This model can be considered as the simplest left-right 
symmetric theory with Majorana neutrinos. This theory provides a simple framework to study lepton number violating processes such as $\mu \to e \gamma$ 
and neutrinoless double-beta decay, which we will study in the future.  

{\textit{Acknowledgments}}: We thank Prof.~Goran Senjanovi\'c for reading the manuscript and very useful comments.

%%%%%%%%%%%%%%%%%%%%%%%%%%%%%

%%%%%%%%%%%%%%%%%%%%

\end{document}